\documentclass[final,3p,times,twocolumn]{elsarticle}

\usepackage[table,xcdraw]{xcolor}
\usepackage{multirow}
\usepackage{hyperref}
\usepackage{amsmath}
\usepackage{siunitx}
\usepackage{graphicx}
\usepackage{lineno}
\journal{Nuclear Instruments and Methods A}

\graphicspath{{./figures/}{./figures/introduction/}{./figures/evaluations/}{./figures/experiment_overview/}{./figures/ratiocalculation/}{./figures/simulation/}{./figures/neutron_energy/}{./figures/energy_shape/}{./figures/isotopics/}{./figures/target_normalization/}{./figures/wraparound/}{./figures/efficiency/}{./figures/uncertainty/}{./figures/validations/}}
\usepackage{caption}
\usepackage{subfigure}

\begin{document}

\begin{frontmatter}

\title{Measurement of material isotopics and atom number ratio with $\alpha$-particle spectroscopy for a NIFFTE fission Time Projection Chamber actinide target.}

\cortext[corresponding]{Corresponding author}

\author[LLNL,corresponding]{M.~Monterial}
\ead{monterial1@llnl.gov}

\author[LANL,ORNL]{K.T.~Schmitt}
\author[LANL]{C.~Prokop}
\author[LANL]{E.~Leal-Cidoncha}
\author[LLNL]{M.~Anastasiou}
\author[LLNL]{N.S.~Bowden}
\author[CSM]{J.~Bundgaard}
\author[LLNL]{R.J.~Casperson}
\author[UC]{D.A.~Cebra}
\author[LLNL]{T.~Classen}
\author[LLNL]{D.H.~Dongwi}
\author[LANL]{N.~Fotiades}
\author[UC,LLNL]{J.~Gearhart}
\author[LANL]{V.~Geppert-Kleinrath}
\author[CSM]{U.~Greife}
\author[LLNL]{C.~Hagmann}
\author[LLNL]{M.~Heffner}
\author[CSM]{D.~Hensle}
\author[CSM,LANL]{D.~Higgins}
\author[ACU]{L.D.~Isenhower}
\author[LLNL]{K.~Kazkaz}
\author[CAL]{A.~Kemnitz}
\author[OSU]{J.~King}
\author[CAL]{J.L.~Klay}
\author[CSM]{J.~Latta}
\author[OSU]{W.~Loveland}
\author[LLNL,AA]{J.A.~Magee}
\author[LANL]{B.~Manning}
\author[LLNL]{M.P.~Mendenhall}
\author[LANL]{S.~Mosby}
\author[LANL]{D.~Neudecker}
\author[LLNL]{S.~Sangiorgio}
\author[LLNL]{B.~Seilhan}
\author[LLNL]{L.~Snyder}
\author[LANL,ANL]{F.~Tovesson}
\author[ACU]{R.S.~Towell}
\author[LLNL]{N.~Walsh}
\author[ACU]{T.S.~Watson}
\author[OSU]{L.~Yao}
\author[LLNL]{W.~Younes}
\author[]{\protect\\(The NIFFTE Collaboration)}

\address[LLNL]{Lawrence Livermore National Laboratory, Livermore, CA 94550, United States}
\address[LANL]{Los Alamos National Laboratory, Los Alamos, NM 87545, United States}
\address[ACU]{Abilene Christian University, Abilene, TX 79699, United States}
\address[CAL]{California Polytechnic State University, San Luis Obispo, CA 93407, United States}
\address[CSM]{Colorado School of Mines, Golden, CO 80401, United States}
\address[OSU]{Oregon State University, Corvallis, OR 97331, United States}
\address[UC]{University of California, Davis, CA 95616, United States}

\fntext[ANL]{Current Address: Argonne National Laboratory, Lemont, IL 60439, United States}
\fntext[ORNL]{Current Address: Oak Ridge National Laboratory, Oak Ridge, TN 37830, United States}
\fntext[AA]{Current Address: Avrio Analytics LLC, Knoxville, TN, 37917 United States}

\begin{abstract}

We present the results of a measurement of isotopic concentrations and atomic number ratio of a double-sided actinide target using $\alpha$-spectroscopy and mass spectrometry. The double-sided actinide target, with predominantly $^{239}$Pu on one side and $^{235}$U on the other, was used in the fission Time Projection Chamber (fissionTPC) for a measurement of the neutron-induced fission cross-section ratio between the two isotopes.  The measured atomic number ratio is needed to extract an absolute measurement fission cross-section ratio. The $^{239}$Pu/$^{235}$U atom number ratio was measured with a combination of mass spectrometry and $\alpha$-spectroscopy with a planar silicon detector achieving uncertainties of less than 1\%. Different strategies for estimating isotopic concentration from the $\alpha$-spectrum are presented to demonstrate the potential of these methods for non-destructive target assay. We found that a combination of fitting spectrum with constraints from mass spectrometry, and summing counts in a region of the spectrum provided the most consistent results with the lowest uncertainty.  
\end{abstract}


\end{frontmatter}


\section{Introduction}
Masses of thin actinide samples, often a mix of several isotopes, are routinely inferred by counting spontaneously emitted $\alpha$-particles with silicon detectors or gas-based 2$\pi$ or 4$\pi$ counters \cite{Pomme2007}. Traditionally, 
when measurements of actinide sample mass have been used to normalize  fission cross-section ratio measurements, specific activity of the sample material was determined using isotopic analysis (e.g. isotopic dilution) and half-lives, while the total mass of the individual isotopes was then determined with $\alpha$ counting~\cite{Meadows1983}. The analysis of the $\alpha$-spectrum was limited to confirming the isotopic composition, rather than an independent measurement of it. In several previous publications of fission cross-section ratios that use this normalization method, many details of the exact methods employed are however omitted~\cite{Staples, Lisowski}.
In this work we provide a detailed account of how the atomic number ratio was determined.  We explore several methods of combining a model $\alpha$-spectrum fit, to estimate specific activity of each isotope directly~\cite{Ihantola2011}, with mass spectrometry used to constrain the fits and determine the isotopic concentrations that are not resolved in the $\alpha$-spectrum.


The Neutron Induced Fission Fragment Tracking Experiment (NIFFTE) collaboration has constructed the fission Time Projection Chamber (fissionTPC) \cite{Heffner2014} with the aim to measure neutron-induced fission cross-section ratios for actinides with total uncertainties better than 1\%.  A high precision measurement of the ratio of $^{239}$Pu and $^{235}$U atoms for the two actinide deposits, referred to hereafter as a target atom number ratio, is needed to provide an absolute $^{239}$Pu(n,f)/$^{235}$U(n,f) fission cross-section ratio \cite{Snyder2021}.

A measurement of the target atom number ratio for a target with deposits of $^{239}$Pu and $^{235}$U, made with a silicon detector, is presented in this article. This measurement was performed with a silicon detector setup, based on the prescription of Pomme~\cite{Pomme2015}, who showed that decay rate measurements with uncertainties as low as 0.1\% are achievable with such devices.  Source-to-detector distances were chosen that strike a compromise between the absolute efficiency and pile-up rate for the two samples. Precise knowledge of absolute efficiency is not necessary for the purpose of measuring the target atom ratio, assuming the ratio of efficiencies for both sides of the target is unity. The sample backing was rotated, so that either the plutonium or uranium faced the detector, without disturbing the source-mount-to-detector distance, as will be described in Section~\ref{sec:silicon}.

\section{Experimental Procedure}
\subsection{Actinide Target}
The target was prepared by depositing $^{239}$Pu and $^{235}$U on a 4~cm diameter and 0.25~mm thick aluminum disk. The actinide deposits were made at the center of the disk, on opposite sides, with a diameter of 2~cm. The uranium sample was deposited using vacuum volatilization, while the plutonium sample was molecular plated~\cite{Loveland,Loveland2016}. The fissionTPC was used to image the target through tracking of the $\alpha$-particles from the decaying actinides back to their origin. 

\begin{figure}[ht]
\begin{center}
\includegraphics[width=1.\linewidth]{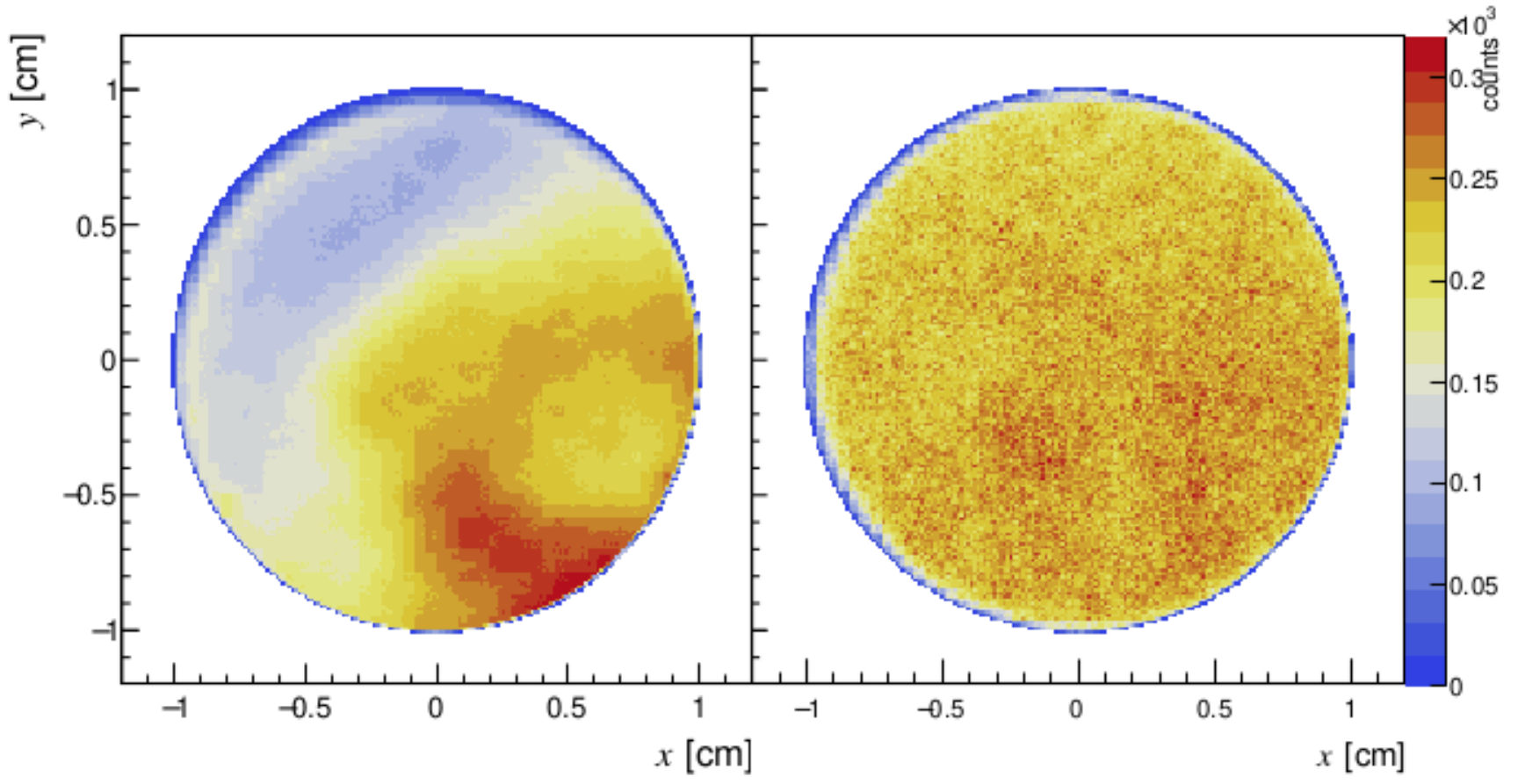}
\end{center}
\caption{Image of $\alpha$-particle track start vertices for plutonium (left) and uranium (right) target deposits.}
\label{fig:vertices}
\end{figure}

The $\alpha$-particle start vertices for each deposit are shown in Fig.~\ref{fig:vertices}.  The pointing resolution of the detector for $\alpha$-particles is approximately 0.3 mm in the plane of the sample. 
The capability to produce vertex images of the target made it also possible to quantify any non-uniformity in the deposit, apparent specifically on the plutonium side. This capability was essential for analyzing the impact of a non-uniform deposit on the detection efficiency of the silicon detector setup, the details of which are presented in Section~\ref{sec:acceptance}

\subsection{Silicon detector setup}
\label{sec:silicon}
\begin{figure}[ht]
\begin{center}$
\begin{array}{cc}
\includegraphics[width=1.\linewidth]{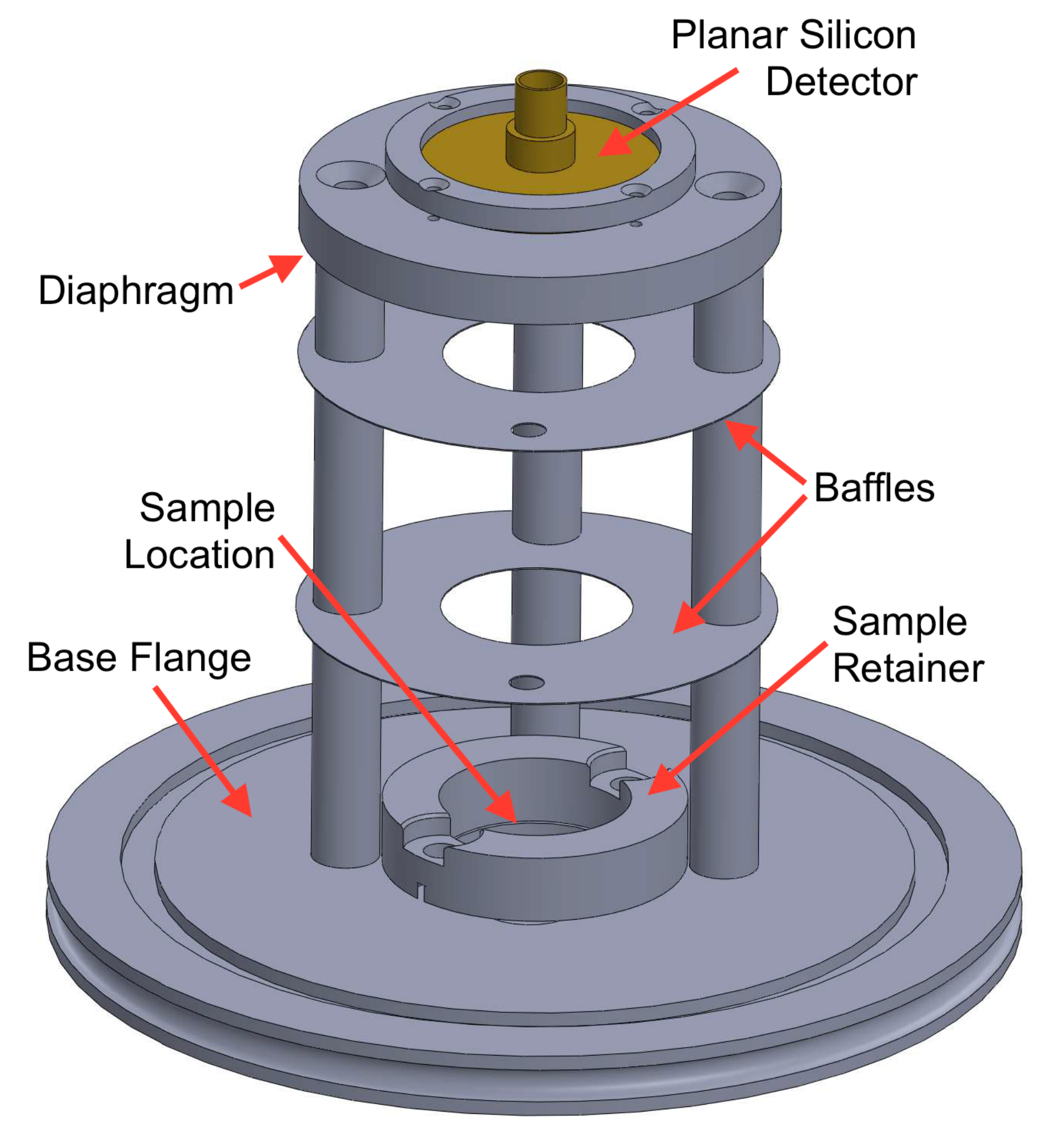}
\end{array}$
\end{center}
\caption{Rendering of the silicon detector setup.  The silicon detector was mounted to an NW160 flange with standoffs.  The plate holding the detector included a diaphragm with an opening diameter of 35.6~mm.  Baffles with inner diameters identical to the diaphragm were mounted between the sample and the detector to block $\alpha$-particles from actinide material that might be sputtered away from the sample.}
\label{fig:auxSetup}
\end{figure}

An ORTEC ULTRA ion-implanted planar silicon detector was used to measure $\alpha$-spectra. The detector had an active area of 1200~mm$^{2}$, minimum depletion depth of 100~$\mu$m and a quoted resolution (FWHM) of 37~keV~\cite{OrtecUltra}.  The sensitive area of the detector was restricted by a stainless steel diaphragm, placed directly in front of the detector, with a thickness of 0.25 mm and an opening diameter of 35.6~mm.  The diaphragm was positioned parallel and concentric to the sample at a distance of 127.5~mm. The fraction of solid angle subtended by the detector to the source was 0.47\%.  Measurements were performed under vacuum, with a nominal absolute pressure of 20 mTorr.  A rendering of the setup is shown in Fig.~\ref{fig:auxSetup}.

A series of precautions were taken to minimize the effect of possible actinide contamination that might be present in the vacuum chamber.  Background data were taken before and after each sample measurement.  Baffles were installed to block $\alpha$-particles originating from outside the target area.  Time under vacuum was minimized to prevent actinide material from leaving the sample.  All background measurements yielded less than one count per four hours in the energy range that would interfere with the target atom ratio measurement.

\subsubsection{Data collection} \label{sec:data_collection}

The actinide samples were measured with the silicon detector in two separate campaigns to validate the work and ensure that the samples were not degraded or otherwise damaged.

In the first measurement campaign collection times were approximately 4 days for uranium and 10 minutes for the plutonium side.  Background data was taken for at least 1 day before loading the sample each time.  In the second measurement campaign uranium data was collected over 56 days, the plutonium in 20 minutes and background in 30 days. Signal processing was performed with a CAEN desktop digitizer, model DT5730, in both campaigns.  

\begin{figure}[ht]
\begin{center}
\includegraphics[width=1.\linewidth]{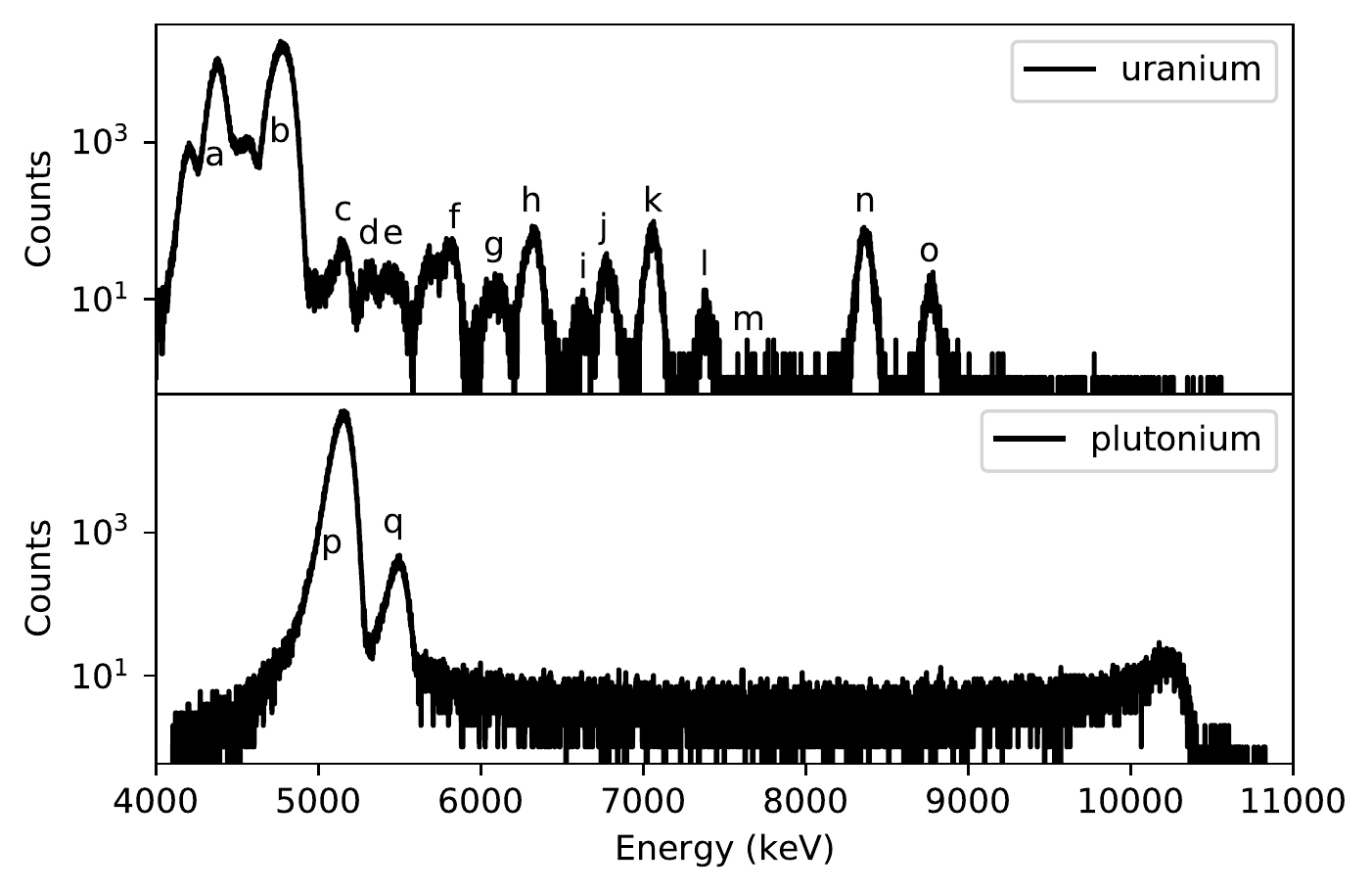}
\end{center}
\caption{Energy spectrum measured by a silicon detector for the uranium deposit (top panel) and plutonium deposit (bottom panel).  The $^{233}$U decay chain is the dominating contributor to counts above 5 MeV on the uranium side.  Peaks are labeled as follows: (a) $^{235}$U, (b) $^{233, 234}$U, (c) $^{239}$Pu, (d) $^{232}$U, (e) $^{228}$Th ($^{232}$U chain), (f) $^{225}$Ac ($^{233}$U chain), (g) $^{221}$Fr ($^{233}$U chain), (h) $^{221}$Fr ($^{233}$U chain), (i) $^{211}$Bi ($^{235}$U chain), (j) $^{216}$Ac ($^{232}$U chain), (k) $^{217}$At ($^{233}$U chain), (l) $^{215}$Po ($^{235}$U chain), (m) $^{214}$Po ($^{234}$U chain), (n) $^{213}$Po ($^{233}$U chain), and (o) $^{212}$Po ($^{232}$U chain).  A peak from $^{238}$Pu (q) was well resolved from $^{239}$Pu and $^{240}$Pu (combined in p) on the plutonium side, and counts from pulse pile-up are evident at energies above 5.6 MeV.}
\label{fig:combinedEnergySpectrum}
\end{figure}

Typical energy spectra are shown in Fig.~\ref{fig:combinedEnergySpectrum}.  Peaks from $^{232}$U, $^{233}$U, $^{234}$U, $^{235}$U, $^{236}$U, $^{239}$Pu background, and $\alpha$-decay daughters are evident in the uranium spectrum. For the plutonium deposit, only three strong peaks were expected from $^{238}$Pu, $^{239}$Pu, and $^{240}$Pu (not resolvable from $^{239}$Pu). 

A series of peaks at energies above 5 MeV was observed for the uranium deposit.  Each peak was associated with a known $\alpha$-particle decay from a daughter of $^{233}$U, $^{234}$U, or $^{235}$U, or direct decays from $^{232}$U or $^{239}$Pu.  Several generations of daughters were observed, ending in the long-lived isotopes of lead.  The decay daughters of $^{233}$U were the largest contributors  to the spectrum at high energy because of the relatively short half life of $^{229}$Th.  The total ratio of daughter and background counts to total counts was measured to be $1.61 \pm 0.02\%$.  The daughter peaks are outside of the analysis region for the uranium side.

\subsubsection{Recovery time}
\label{subsec:dead_time}

The observed event rates were around 0.58 counts per second for uranium, and 3460 counts per second for plutonium. The total background rates were 0.01 counts per second. The total recovery time is on the order of microseconds, therefore it is only of consequence in the higher rate plutonium measurement. 

The total recovery time is the time it takes to detect two distinct pulses, and is the result of overlapping effects: shaping time, pile-up, detector dead time, and the digitizer acquisition window. As a result of these effects the observed count rate is lower than the true count rate.

The true count rate can be estimated from the time distribution of observed events, since we expect these to follow a Poisson random process. The distribution of the time differences, $t$, between these events follows an exponential 

\begin{align} 
	f(t) = m C \exp(-Ct) \label{eq:time_dist}
\end{align}
where $m$ is the total number of counts and $C$ is the true count rate. The estimated true count rate from the fit to the distribution of events in the plutonium measurement was 3504$\pm$2 counts per second. Therefore, the true count rate in the plutonium measurement is 1.1\% higher than the observed count rate. The recovery time was calculated from 
\begin{align}
\tau = \frac{(C - N)}{C \cdot N}
\end{align}
where $N$ is the observed count rate. The total recovery time was estimated to be 3.59$\pm$0.2 $\mu s$.

\subsubsection{Absolute Efficiency Ratio}
\label{sec:acceptance}

The actinide deposits were both 2~cm in diameter and arranged back-to-back on an aluminum disk.  The detector geometry was designed such that the the disk could be rotated with either deposit facing the silicon detector so that the source-to-detector distance was unchanged and therefore the absolute detector efficiency for each actinide source would be identical. The absolute efficiency of the silicon detector is defined as
\begin{align}
\epsilon= \frac{n}{N} \,\text{,}
\end{align}
where $n$ is the number of $\alpha$'s detected by the Si detector and $N$ is the total number of $\alpha$'s generated by the deposit.

The absolute detector efficiency is subject to two potential sources of systematic uncertainty.  The first source of uncertainty is a result of the different distribution of target material for the $^{239}$Pu and  $^{235}$U, as shown in Fig.~\ref{fig:vertices}.  The second source of systematic uncertainty arises from clearance between the aluminum disk and the base flange. This clearance is 0.32 mm and could result in possible relative position offsets between the $^{239}$Pu and $^{235}$U targets, as the respective targets need not be sitting flush at the origin $(0,0,0)$.

A Geant4 simulation of the Si detector system was developed in order to perform a systematic study of the efficiency ratio. In the simulation, $\alpha$'s were considered as the primary particle with initial energies assigned from $\alpha$-decay energies of $^{239}$Pu and $^{235}$U, respectively. Starting $(x_i,y_i)$ vertex distributions of both targets were taken from the fissionTPC data of Fig.~\ref{fig:vertices}. Simulations were performed by shifting the center of the Al disk over a range of $0-0.3$ mm in the $x-y$ plane. Since the $^{239}$Pu target was non-uniform, negative displacements of the Al disk were also considered.  The $z$ position was considered to be fixed for both $^{239}$Pu and $^{235}$U targets since gravity will hold the target flush with the base flange.

Calculation of the total efficiency ratio 
\begin{align}
R_\epsilon= \frac{\epsilon_{Pu(x_i,y_i)}}{\epsilon_{U(x_i,y_i)}}
\end{align}
was carried out by taking the average of the permutations of the aforementioned relative efficiency ratios of $^{239}$Pu and $^{235}$U. In particular, this process involved evaluating combinatorial efficiency ratios for each assumed $i^{th}$ $^{239}$Pu start vertex $(x_i,y_i)$ over the entire suite of assumed $^{235}$U $(x_i,y_i)$ start vertices. The average of this combinatorial sum was taken to be the efficiency ratio $R_{\epsilon}=1.00006\pm5.54\times10^{-5}$, which is consistent with unity at the sub-percent level.  We therefore consider the absolute detector efficiency to be equal for each actinide measurement and any contribution to the atom number ratio uncertainty to be negligible.



\section{Results}
\label{sec:results}

\subsection{Mass Spectrometry}

Mass spectrometry of surrogate samples was used to estimate isotopic concentrations in the targets, some of which were not resolvable in the $\alpha$-spectrum. Two samples were analyzed via mass spectrometry for uranium (\#1 and \#2) and two samples were analyzed for plutonium (\#3 and \#4). 
The mass spectrometry results are shown in Table~\ref{table:Isotopics}.
In each case one sample (\#1 and \#3) was raw material used to make the target, and the other sample (\#2 and \#4) was a deposit prepared in a similar way to the target under study. The mass spectrometry uncertainties were determined as described in Ref.~\cite{RossUncert}. Published values for half-lives were used to calculate the decay constants and are shown in Table~\ref{table:halflivesDDEP} ~\cite{DDEP,NNDC}. The $\alpha$ energies and intensities are shown in Table~\ref{table:alpha_energies}.

\begin{table*}[ht] 
	\begin{center}
		\caption{Uranium and plutonium target isotopic atom percentages as measured with mass spectrometry\cite{Ross}}
		\label{table:Isotopics}
		\begin{tabular}{ l S[table-format=3.6] S[table-format=3.6] S[table-format=3.6]  S[table-format=3.6]  }
			\hline
			\bf{Isotope} & {Sample \#1} & {Uncertainty} & {Sample \#2} & {Uncertainty} \\
			\hline
			$^{233}$U & .01886 & .00004 & .01893 & .00008  \\
			$^{234}$U & .03448 & .00032 & .03536 & .00004 \\
			$^{235}$U & 99.677 & .002 & 99.634 & .014 \\ 
			$^{236}$U & .1701 & .00177 & .1763 & .0005 \\
			$^{238}$U & .0998 & .0005 & .1355 & .0014 \\
			\hline

  \bf{Isotope} & {Sample \#3}  & {Uncertainty} & {Sample \#4}  & {Uncertainty}  \\
  \hline 
  $^{238}$Pu & \multicolumn{4}{c}{not included in analysis} \\
  $^{239}$Pu & 99.1323 & 0.0024 & 99.1213 & 0.0008  \\
  $^{240}$Pu & 0.8675 & 0.0023 & 0.8770 & 0.0008 \\
  $^{241}$Pu & \multicolumn{2}{c}{not detected} & 0.001427 & 0.000010 \\ 
  $^{242}$Pu & 0.000242 & 0.000042 & 0.000250 & 0.000006 \\
  \hline
\end{tabular}
\end{center}
\end{table*}
\begin{table}[ht]
\begin{center}
\caption{Half-lives for $\alpha$~emitters in target, taken from DDEP~\cite{DDEP}.  Note that since no decay data were available from DDEP for $^{233}$U, the half-life reported by the NNDC~\cite{NNDC} was used for that isotope.}
\label{table:halflivesDDEP}
\begin{tabular}{l S S[table-format=1.10e1]}
  \hline
  Isotope & {half-life (s)} & {Uncertainty (s)} \\
  \hline
  $^{233}$U \cite{NNDC}  & \num{5.0240e12}   & \num{6.3e9}  \\
  $^{234}$U \cite{DDEP}  & \num{7.747e12}    & \num{1.9e10} \\
  $^{235}$U \cite{DDEP}  & \num{2.2217e16}   & \num{3.2e13}  \\
  $^{236}$U \cite{DDEP}  & \num{7.394e14}    & \num{1.9e12} \\
  $^{238}$U \cite{DDEP}  & \num{1.4100e17}  & \num{1.6e14}  \\
  $^{238}$Pu \cite{DDEP} & \num{2.76886e9}   & \num{9.5e5}   \\
  $^{239}$Pu \cite{DDEP} & \num{7.6054e11}   & \num{3.5e8}  \\
  $^{240}$Pu \cite{DDEP} & \num{2.0705e11}   & \num{2.2e8}   \\
  \hline
\end{tabular}
\end{center}
\end{table}

\begin{table}[h]
	\caption{Energies and relative intensities of $\alpha$ particles from isotopes included in the fitted alpha spectrum, taken from DDEP~\cite{DDEP}.  Only intensities greater than 1\% are shown.}
	\begin{center}
	\label{table:alpha_energies}
	\begin{tabular}{c | c | c}
		\hline
		{Isotope} & {Energy (keV)} & Intensity (\%) \\ \hline
		\multirow{3}{*}{ $^{233}$U}  & 4729.0                            & 1.61      \\
		& 4783.5                            & 13.20          \\
		& 4824.2                            & 84.30                   \\ \hline
		\multirow{2}{*}{$ ^{234}$U}        & 4722.4                            & 28.42          \\
		& 4774.6                            & 71.37          \\ \hline
		\multirow{8}{*}{$ ^{235} $U}        & 4214.7                            & 5.95           \\
		& 4322.0                            & 3.33           \\
		& 4366.1                            & 18.80          \\
		& 4397.8                            & 57.19          \\
		& 4414.9                            & 3.01           \\
		& 4502.4                            & 1.28           \\
		& 4556.0                            & 3.79           \\
		& 4596.4                            & 4.74           \\ \hline
		\multirow{2}{*}{$^{236}$U}        & 4445.0                            & 26.10          \\
		& 4494.0                            & 73.80          \\ \hline
		\multirow{2}{*}{$^{226}$Ra}       & 4601.0                            & 5.95           \\
		& 4784.3                            & 94.04          \\ \hline
		\multirow{2}{*}{$^{238}$Pu}       & 5456.3                            & 28.85          \\
		& 5499.0                            & 71.04          \\ \hline
		\multirow{3}{*}{$^{239}$Pu}       & 5105.8                            & 11.87          \\
		& 5143.8                            & 17.14          \\
		& 5156.6                            & 70.79          \\ \hline
		\multirow{2}{*}{$^{240}$Pu}       & 5123.6                            & 27.16          \\
		& 5168.1                            & 72.74          \\ \hline
	\end{tabular}
	\end{center}
\end{table}

For the uranium case, the differences between the measured isotopic ratios are consistent with the addition of a small amount of natural uranium in processing from the raw material (Sample \#1) to the target deposit via vacuum volatilization (Sample \#2).  This difference is demonstrated most clearly in the case of $^{238}$U, which has the largest relative discrepancy and is the majority constituent of natural uranium. The decay fraction for  $^{235}$U obtained from the $\alpha$-spectrum was consistent with Sample \#2. 

The $^{239}$Pu decay fraction is not highly sensitive to measured isotopic ratios for plutonium.  The $^{238}$Pu was not resolvable from $^{238}$U background in the mass spectrometry measurements, but it was measured with $\alpha$-spectrometry from the silicon detector.  The difference between calculated decay fractions using the two sets of ratio data is 0.03\%.  The uncertainty on the decay fraction for $^{239}$Pu, 0.10\%, is dominated by uncertainties in the published half-lives, so if we assume that the variation in isotopic abundances for the target under study is within a factor of ten of the two mass spectrometry samples, this contribution to the uncertainty is negligible.

\subsection{Silicon Detector Data Analysis}
\label{subsec:analysis}

\subsubsection{Uranium Sample} \label{sec:uranium_sample}

The full $\alpha$-spectrum was modelled by assigning an exponentially modified Gaussian to each peak and summing over the decay fractions, $d$, and alpha emission intensities, $I_{\alpha}$, as
\begin{align}
S =& \sum_d \sum_{I_{\alpha}} d I_{\alpha}  f_{EMG}(\mu, \sigma, \omega)  \label{eq:spectrum_model} \\
\mu =& E_{\alpha} c_1 + c_0
\end{align}
where $\mu$ is the peak mean location with energy $E_{\alpha}$ and calibration constants $c_1$ and $c_0$, and $\sigma$ and $\omega$ are the standard deviation and skewness parameters. The calibration constants were determined by letting all the parameters float. Those constants were then fixed for all subsequent fits.  Only alpha emission lines with relative intensities greater than 0.1\% were included. The contribution from $^{238}$U was not included because it is insignificant. 

\begin{figure}[ht]
	\begin{center}$
		\begin{array}{cc}
		\includegraphics[width=1.\linewidth]{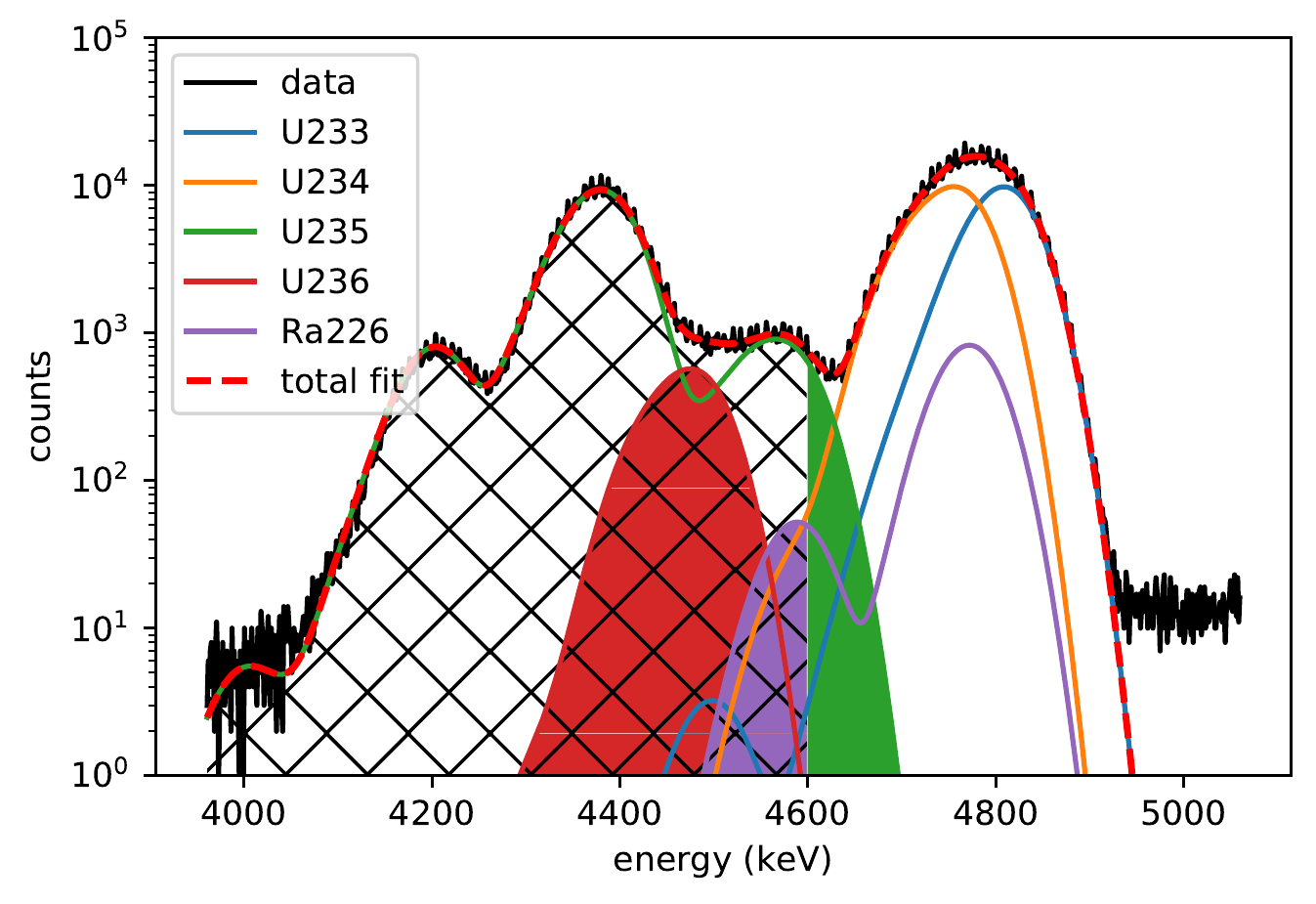}
		\end{array}$
	\end{center}
	\caption{The fits to the uranium $\alpha$-spectrum from the first measurement campaign, with $^{236}$U  fixed and $^{226}$Ra included . The region of interest is shown in hashes between 3960 keV and 4600 keV. The partial integral of $^{235}$U contribution was added to the counts in the ROI, and the counts from other isotopes inside the ROI were subtracted. These adjustments are shown as solid regions under the fits.}
	\label{fig:u_spectrum_fit}
\end{figure}

Two methods were considered when calculating the contribution of $^{235}$U to the $\alpha$-spectrum. In the first method the integrals of the fits were used directly as an estimate of the $^{235}$U contribution (Fit-Only). In the second method, a region of interest (ROI) was selected over the portion of the spectrum with majority $^{235}$U contribution. The fit results from other isotopes in the ROI were subtracted, and $^{235}$U fit results outside the ROI were added back into the estimate. The ROI method has the advantage of having lower uncertainty due to a smaller contribution of errors from the fit. The results of the ROI method are shown in Fig.~\ref{fig:u_spectrum_fit}. 

In addition, we considered constraining the relative contribution of each uranium isotope from the results of mass spectrometry shown in Table~\ref{table:Isotopics}. We tested fixing all contributions (All fixed), fixing just the $^{236}$U contribution ($^{236}$U fixed), and letting all the contributions float (All floating). The last method does not require the results of mass spectrometry, which are higher fidelity but were performed on surrogate samples. We settled on using the middle approach of fixing $^{236}$U, since its contribution is unresolvable from $^{235}$U. 

The last unresolved factor was the unknown contribution of $^{226}$Ra, the presence of which is indicated by the existence of resolvable uranium decay daughters in the $^{226}$Ra decay chain.
We tested the proposition of including it along with the other isotopes of uranium in the spectrum model from Eq.~\ref{eq:spectrum_model}. The results from all these approaches are shown in Table \ref{tab:uranium_fits_table}. Although the contribution of $^{226}$Ra did not have significant impact on the result with either method, we included it in our estimates for completeness since it met the criterion of $>$0.1\% relative intensity. 

The Fit-Only approach resulted in a consistently lower estimate of the number of $^{235}$U atoms, except for the case with the contributions fixed by mass spectrometry results (All fixed).  The Fit-Only method relies more on the ability of Eq.~\ref{eq:spectrum_model} to model the data, and by extension it is impacted by any incompleteness or inaccuracies in the accepted $\alpha$ emission intensity or any behavior of the detector that is not well modeled by an exponentially modified Gaussian. 
By contrast, the ROI method is more consistent, and it produces a lower overall uncertainty in the estimate of the number of $^{235}$U atoms. Based on these results, the ROI method with $^{226}$Ra included and fixed $^{236}$U contribution was chosen as the method of determining the target atom number ratio presented in this paper and used for the fissionTPC cross-section ratio~\cite{Snyder2021}.

\begin{table*}[ht]
	\caption{Summary of the various approaches to estimating the number of atoms of $^{235}$U from the $\alpha$-spectrum of the second measurement campaign. Based on these results, the ROI method with $^{226}$Ra included and fixed $^{236}$U contribution was chosen as the method of determining the target atom number ratio presented in this paper and used for the fissionTPC cross-section ratio~\cite{Snyder2021}.}
	\label{tab:uranium_fits_table}
    \centering
	\begin{tabular}{|c|c|c|c|c|c|}
		\hline
		&                              & \multicolumn{3}{c|}{\textbf{Constraints}}                                         &                                            \\ \cline{3-5}
		\multirow{-2}{*}{\textbf{\begin{tabular}[c]{@{}c@{}}Integration\\ Method\end{tabular}}} & \multirow{-2}{*}{$^{226}$Ra} & All floating              & $^{236}$U fixed           & All fixed                 & \multirow{-2}{*}{\textbf{$^{235}$U Atoms}} \\ \hline
		& \cellcolor[HTML]{C0C0C0}     & \cellcolor[HTML]{C0C0C0}x & \cellcolor[HTML]{C0C0C0}  & \cellcolor[HTML]{C0C0C0}  & \cellcolor[HTML]{C0C0C0}8.61 $\pm$ 0.06    \\ \cline{2-6} 
		& x                            & x                         &                           &                           & 8.59 $\pm$ 0.05                            \\ \cline{2-6} 
		& \cellcolor[HTML]{C0C0C0}     & \cellcolor[HTML]{C0C0C0}  & \cellcolor[HTML]{C0C0C0}x & \cellcolor[HTML]{C0C0C0}  & \cellcolor[HTML]{C0C0C0}8.63 $\pm$ 0.05    \\ \cline{2-6} 
		& x                            &                           & x                         &                           & 8.60 $\pm$ 0.05                            \\ \cline{2-6} 
		\multirow{-5}{*}{\textbf{Fit-Only}}                                                     & \cellcolor[HTML]{C0C0C0}     & \cellcolor[HTML]{C0C0C0}  & \cellcolor[HTML]{C0C0C0}  & \cellcolor[HTML]{C0C0C0}x & \cellcolor[HTML]{C0C0C0}8.69 $\pm$ 0.02    \\ \hline
		&                              & x                         &                           &                           & 8.69 $\pm$ 0.03                            \\ \cline{2-6} 
		& \cellcolor[HTML]{C0C0C0}x    & \cellcolor[HTML]{C0C0C0}x & \cellcolor[HTML]{C0C0C0}  & \cellcolor[HTML]{C0C0C0}  & \cellcolor[HTML]{C0C0C0}8.67 $\pm$ 0.03    \\ \cline{2-6} 
		&                              &                           & x                         &                           & 8.70 $\pm$ 0.02                            \\ \cline{2-6} 
		& \cellcolor[HTML]{C0C0C0}x    & \cellcolor[HTML]{C0C0C0}  & \cellcolor[HTML]{C0C0C0}x & \cellcolor[HTML]{C0C0C0}  & \cellcolor[HTML]{C0C0C0}8.68 $\pm$ 0.02    \\ \cline{2-6} 
		\multirow{-5}{*}{\textbf{ROI}}                                                          &                              &                           &                           & x                         & 8.70 $\pm$ 0.02                            \\ \hline
	\end{tabular}
\end{table*}

\subsubsection{Plutonium Sample}

The high count rate from the $^{239}$Pu side of the sample necessitates additional analysis prior to fitting.  The waveform from each event was analyzed for pulse height and pulse area, and pileup was rejected event-by-event via comparison of these extracted quantities.  All events that occurred within 10 microseconds of each other were rejected and the total number of counts was determined using Eq. \ref{eq:time_dist} fit to the data. The remaining events were histogrammed and fit with a linear combination of Eq. \ref{eq:spectrum_model} comprised of every known alpha transition for $^{238}$Pu,$^{239}$Pu,$^{240}$Pu, and $^{242}$Pu.  The alpha decays from $^{239}$Pu and$^{240}$Pu are not resolvable with the current resolution of our system so they were constrained in the fit as a ratio to $^{239}$Pu using the mass spectrometry results presented in Table \ref{table:decay_fractions}.  The contribution from $^{238}$Pu was resolved and as such was left unconstrained in the fit.  The ROI technique was employed in a similar fashion as described for the $^{235}$U in Section~\ref{sec:uranium_sample}, with a boundary located where the low-energy tail of $^{238}$Pu is of equal intensity to the high-energy tail of the remaining unresolved plutonium spectra.  The results of the $^{238}$Pu fit was subtracted from the low-energy ROI and added to the high-energy ROI.  The converse was performed for the remaining plutonium isotopes. 

\begin{table}[h] 
	\begin{center}
		\caption{Decay fraction estimates averaged between the two measurements.}
		\label{table:decay_fractions}
		\begin{tabular}{l S S }
			\hline
			Isotope & {Decay fraction} & {Uncertainty (\%)}\\
			\hline
			$^{233}$U  &  0.289 & 0.60 \\
			$^{234}$U  & 0.349 & 0.49\\
			$^{235}$U  & 0.343 & 0.41 \\
			$^{236}$U  &  0.0184 & 0.35 \\
			$^{238}$Pu &  0.0083 & 5.9 \\
			$^{239}$Pu &  0.9607 & 0.19\\
			$^{240}$Pu &  0.03209 & 0.22 \\
			$^{242}$Pu &  1.5e-7 & 8.7 \\
			\hline
		\end{tabular}
	\end{center}
\end{table}

\subsubsection{Target Atom Ratio}

The target atom ratio was calculated separately for both of the silicon detector measurement campaigns, introduced in Section~\ref{sec:silicon} and the results are shown in Table~\ref{table:ratioResults}. A summary of the contributing factors to the uncertainty in the total number of atoms for each isotope for the second measurement is shown in Table~\ref{table:siUncertainties}. 

The two measurement campaigns were conducted independently by two different teams using the same instrument, and were in agreement.  We combined the partial uncertainties from both measurements assuming they are correlated, with the exception of counting statistics which we added in quadrature. The individual measurement uncertainties shown in Table~\ref{table:ratioResults}. The average target atom ratio is 1.7343$\pm$0.0050 (0.288\%).

\begin{table}[ht]
\caption{Target atom number ratio results and uncertainties for the two measurement campaigns. }
\label{table:ratioResults}
\begin{center}
\begin{tabular}{lcc}
  \hline
   &  Meas. 1  & Meas. 2 \\
  \hline
  $^{235}$U atoms 				& $8.647 \times 10^{17}$ & $8.682 \times 10^{17}$ \\  
  $^{235}$U Uncertainty 	&  0.257\% 						& 0.187\% \\
  $^{239}$Pu atoms 			& $4.996 \times 10^{17}$ & $5.00 \times 10^{17}$ \\
  $^{239}$Pu Uncertainty & 0.214\% 							& 0.235\% \\
  Atom number ratio 		& 1.731 							& 1.736 \\
  Total Uncertainty 			& 0.334\% 						& 0.30\% \\
  \hline
\end{tabular}
\end{center}
\end{table}

\begin{table}[ht]
	\caption{Uncertainty contributions for the target atom numbers from the second measurement campaign.}
	\label{table:siUncertainties}
	\begin{tabular}{lcc}
		\hline
		Source of Uncertainty & $^{235}$U & $^{239}$Pu  \\
		\hline
		Spectrum fits error  &  0.058\%  &  0.007\% \\
		Counting statistics  &  0.102\%   & 0.078\%  \\
		Mass spectrometry &   0.019\%  &   0.17\%     \\
		Recovery time          & n/a          & 0.07\% \\
		Half life                   & 0.144\%   & 0.124\% \\
		\hline
		Total                      & 0.187\%     & 0.235\% \\
		\hline
	\end{tabular} 
\end{table}

\section{Conclusions}
A complementary and independent measurement of the atom number ratio for the targets used in the NIFFTE collaboration’s fissionTPC measurement of the $^{239}$Pu(n,f)/$^{235}$U(n,f) cross-section ratio have been performed using a combination of mass spectrometry and alpha spectroscopy with a silicon detector. The two silicon detector measurement campaigns were conducted independently by two different teams using the same instrument. The target atom ratio for the two measurements was in agreement within the reported uncertainties. The average target atom ratio is 1.7343$\pm$0.0050 (0.288\%). For a complete description of the results of the fissionTPC measurement of the normalized $^{239}$Pu(n,f)/$^{235}$U(n,f) cross-section ratio and comparisons to other measurements see \cite{Snyder2021}.

The methods described in this paper improve upon previous efforts to normalize fission cross-section ratio measurements by utilizing fitting of the $\alpha$-spectrum to determine the isotopic concentrations of our samples. While it was possible to use just $\alpha$-spectrum analysis to determine these quantities, we found that combining this analysis with constraints from mass spectrometry resulted in more consistent results. Uncertainty in the $\alpha$ decay data and the robustness of the $\alpha$-spectrum model ultimately limit the precision of this technique, but it provides an alternative when destructive methods like mass spectrometry are not possible.


\section{Acknowledgements}
This work was funded by the U.S. Department of Energy and operated by Los Alamos National Security, LLC, under contract DE-AC52-06NA25396. This work performed under the auspices of the U.S Department of Energy by Lawrence Livermore National Laboratory under contract DE-AC52-07NA27344. This material is based upon work supported by the U.S. Department of Energy, National Nuclear Security Administration, Stewardship Science Academic Alliances Program, under Award Number DE-NA0002921.

LLNL-JRNL-820554


\bibliography{targetCounting}

\end{document}